\documentclass[aps,prb,amsmath,amssymb,twocolumn,notitlepage,nobibnotes,nolongbibliography]{revtex4-2}

\usepackage{graphicx,graphics,color,epsfig,exscale}
\usepackage{bm}
\usepackage{bbm}
\usepackage{float}
\usepackage{mathrsfs}
\usepackage[utf8]{inputenc}
\usepackage{pdfsync}
\usepackage[colorlinks=true,bookmarks=true,pdfborder={0 0 0},linkcolor=blue,urlcolor=blue,breaklinks=true,bookmarksnumbered=true]{hyperref}
\usepackage{xcolor}
\usepackage[makeroom]{cancel}
\usepackage{subfigure}
\usepackage{times}
\usepackage{feyn}
\def \x{\bm{x}}

\def \k{\bm{k}}

\def \i{\text{\bf{i}}}

\def \vQ{\vec{Q}}

\def \hc{\hat{c}}

\def \hn{\hat{n}}

\def \hs{\hat{s}}

\def \hu{\hat{u}}
\def \hv{\hat{v}}

\def \hD{\hat{D}}

\def \hO{\hat{O}}

\def \heta{\hat{\eta}}

\def \hrho{\hat{\rho}}

\def \mT{\mathcal{T}}
\def \mH{\mathcal{H}}

\def \pt{\partial}
\def \si{\sigma}
\def \sib{\bar{\sigma}}

\def \sgn{\text{sgn}}
\def \Imm{\mathfrak{Im}}

\def \tr{\text{Tr}}
\def \Xs0{X^{\sigma 0}}
\def \X0s{X^{0 \sigma}}

\newcommand{\abs}[1]{\lvert#1\rvert}

\newcommand{\ev}[1]{\mbox{$\langle #1 \rangle$}}
\newcommand{\dev}[1]{\mbox{$\langle\langle #1 \rangle\rangle$}}

\newcommand{\ket}[1]{\mbox{$| #1 \rangle$}}

\newcommand{\ua}{\uparrow}
\newcommand{\da}{\downarrow}

\newcommand{\iden}{ \mathbbm{ 1}}

\title{Algebraic-Dynamical Perturbation Theory of Large-\(U\) Hubbard Models. Single Particle Spectrum of Antiferromagnetic Mott Insulating States.}
\hypersetup{
 pdfauthor={Wenxin Ding},
 pdftitle={Algebraic-Dynamical Perturbation Theory of Large-\(U\) Hubbard Models. Single Particle Spectrum of Antiferromagnetic Mott Insulating States.},
 pdfkeywords={},
 pdfsubject={},
 pdfcreator={Emacs 31.0.50 (Org mode 9.8-pre)},
 pdflang={English}}
\begin{document}

\title{Algebraic-Dynamical Perturbation Theory of Large-\(U\) Hubbard Models. Single Particle Spectrum of Antiferromagnetic Mott Insulating States.}
 \date{\today} \author{Wenxin Ding$^{1}$} \thanks{wxding@ahu.edu.cn} \author{Rong Yu$^{2,3}$} \affiliation{$^1$School of Physics, Anhui University, Hefei, Anhui Province, 230601, China  \\  $^2$Department of Physics and Beijing Key Laboratory of Opto-electronic Functional Materials \& Micro-nano Devices, Renmin University of China, Beijing 100872, China \\ $^3$Key Laboratory of Quantum State Construction and Manipulation (Ministry of Education), Renmin University of China, Beijing, 100872, China}
  \begin{abstract} In this work, we present an analytical framework for studying antiferromagnetic (AFM) Mott insulating states in the Hubbard model. We first derive an analytical solution for the single-particle Green's functions in the atomic limit. Within a second-order perturbation approach, we compute the ground state energy and show that the ground state is antiferromagnetically ordered. Then we derive an analytical solution for single-particle Green's functions when effects of the hopping term are considered in the N\'{e}el state. \textcolor{black}{With the analytical solution, we compute the spectral functions and explain various properties of the AFM Mott insulating state as} observed both experimentally and numerically: i) magnetic blueshift of the Mott gap; ii)  \textcolor{black}{the low energy part in the} parental compounds of cuprate high \(T_c\) superconductors  \textcolor{black}{which corresponds to a single band Hubbard model description.} This work comprehends the electronic properties of antiferromagnetic Mott states analytically and provides a foundation for future investigations of doped antiferromagnetic Mott insulators, aiming for the mechanism of cuprates high-\(T_c\) superconductivity.\end{abstract}
 \maketitle
\section{Introduction}
\label{sec:org0686bda}
The mechanism of high \(T_c\) superconductivity in cuprates\cite{bednorz-1986-possib-hight} is one of the most challenging problems in modern condensed matter physics. Although they are generally three-band, charge transfer insulators\cite{zaanen-1985-band-gaps}, it is widely accepted that the mechanism is rooted in the physics of doped antiferromagnetic (AFM) Mott insulators\cite{lee-2006-dopin-mott-insul}, which can be adequately described by single band Hubbard models (HM)\cite{hubbard-1963-elect-correl,arovas-2022-hubbar-model} with a large onsite repulsion \(U\).

However, despite its seemingly simplicity, we still do not have a well-controlled and comprehensive analytic theory for the strong coupling limit of the HM, except for some aspects of its properties. For example, the AFM order at half-filling was long established through the superexchange mechanism \cite{anderson-1950-antif}. But incorporation of antiferromagnetism within a systematic solution of HM on equal footing with charge dynamics, often through slave particle constructions\cite{kotliar-1986-new-funct,yoshioka-1989-slave-fermion,han-2019-finit-temper,ding-2019-effec-exchan,florens-2002-quant-impur}, appears to be difficult and remains unsolved to this day.
\textcolor{black}{This causes difficulty in understanding various important spectral features of the AFM Mott insulating state observed in experiments and/or numerics, including the magnetic blueshift of the Mott gap and the waterfall-like band in parent compounds of cuprate superconductors. Lacking a proper theory for AFM Mott states also hinders further study for the doped cases, which is believed to be relevant to high-temperature superconductivity.}
In fact, even theoretical understanding for single\cite{ye-2013-visual-atomic,ding-2024-local-densit} or a few impurities\cite{cai-2016-visual-evolut} proves to be difficult in the absence of such a theory.

The common difficulties encountered in previous works are i) noncanonical operator algebras and ii) lack of Wick's theorem. Both make perturbation sequence difficult to control. Recently, Shastry group pioneered a series of works studying the \(t-J\) model\cite{zhang-1988-effec-hamil}, the \(\text{{\it extremely correlated Fermi liquid}}\) (ECFL) theory\cite{shastry-2011-extrem-correl,shastry-2013-extrem-correl}, and obtained results quantitatively comparable to numerical methods such as dynamical mean field theories\cite{zitko-2013-extrem-correl} and density matrix renormalization group\cite{mai-2018-t-t}. ECFL overcomes these difficulties by combining the noncanonical algebras with Heisenberg-equations-of-motion (HEOM) to derive a close-form Schwinger-Dyson-equations-of-motion (SDEOM) for the dynamical Green's functions (GFs). W. Ding further generalized the ECFL idea to arbitrary operators and many-body correlations which is called the \emph{algebraic-dynamical theory} (ADT) \cite{Ding-2022-algeb-dynam}.

{\color{black}
Historically, many versions of slave particle constructions were proposed and applied to study HMs and other models of strong correlations. While these studies made significant progresses in some aspects, it is desirable to unite these different constructions since the underlying model is the same. For all these constructions, a successful description of a doped AFM state is still not achieved. In fact, even for the simplest half filling case, although people have succeeded in describing the Mott gap related physics with dynamical slave bosons, such as slave rotor\cite{florens-2002-quant-impur} and slave spin\cite{ding-2024-dynam-t,demedici-2005-orbit-selec}, the incorporation of antiferromagnetism on equal footing via such constructions still remains unsolved\cite{Ding-2019-effe-exch}. This is due to the artificial spin-charge separation introduced by fractionalizing the physical electron operators which carry both spin and charge. The ADT method not only provides a comprehensive understanding in all slave-particle methods, but its construction also naturally accounts for spin-charge coupling by working with the original physical operators and their exact non-commutative algebras.
}

In this work, we develop a self-contained and analytic treatment of the HM in the large-\(U\) limit at half-filling. We first solve the atomic HM at a single site exactly. With the atomic limit solution, we perform a second-order dynamical and perturbative calculation of the ground state energy to show that the lattice ground state is antiferromagnetically ordered. Then we derive an analytic solution for single-particle GFs in such AFM Mott states of the HM on a square lattice with only nearest neighbor (nn) hopping. Finally, we study the physical properties through the single-particle GFs and compare with previous numerical studies and experimental observations.
\textcolor{black}{Our results successfully captures several key features of the AFM Mott insulating state, including the magnetic blueshift of the Mott gap and the high-energy waterfull-like band.}
Through such comparison, we demonstrate that our analytic theory provides an accurate and efficient description for the AFM Mott insulating states of the HM.
\section{Algebraic-Dynamical Theory: A Brief Review}
\label{sec:orgb1660de}
In this work, we study the HM\cite{hubbard-1963-elect-correl} in the large\(-U\) limit {\color{black} with the ADT method developed by one of the authors in Ref. \cite{Ding-2022-algeb-dynam}. In this section, we briefly review the computation framework of ADT.
}
\subsection{Complete Operator Basis Sets}
\label{sec:org3ab80c9}
ADT employs \({\text{\it complete operator basis sets}}\) (COBSs) \cite{schwinger-1960-unitar-operat-bases,fano-1957-descr-states} to formulate a complete description of interacting quantum systems. A COBS is a complete collection of operators \(\mathscr{U} = \{\hu^\alpha\}\) of the system. The \emph{completeness} requires elements of a COBS satisfy
\begin{align}
\text{orthogonality condition:}\quad & \tr ((\hu^\alpha)^\dagger \hu^\beta) = C ~\delta_{\alpha \beta}, \label{eq:d1}\\
\text{a set of {\it closed} algebras:}\quad & \hu^\alpha \hu^\beta = \sum_{\gamma} a^{\alpha \beta}_\gamma \hu^{\gamma}. \label{eq:d2}
\end{align}
The trace operation \(\tr ((\hu^\alpha)^\dagger \hu^\beta) = (\hu^\alpha, \hu^\beta)\) can be viewed as an inner product and \(C\) is typically chosen to be 1. The closed algebra can be decomposed into a symmetric (bosonic) sector \([\hu^\alpha, \hu^\beta ] = \sum_{\gamma} b^{\alpha \beta}_\gamma \hu^\gamma\) and anti-symmetric (fermionic) sector \textasciitilde{} \(\{ \hu^\alpha, \hu^\beta \} = \sum_{\gamma} f^{\alpha \beta}_\gamma \hu^\gamma\), where \(b\)s and \(f\)s are structure constants relating to \(a's\) by \(b^{\alpha \beta}_\gamma = a^{\alpha \beta}_\gamma - a^{\beta \alpha}_\gamma, \quad f^{\alpha \beta}_\gamma = a^{\alpha \beta}_\gamma + a^{\beta \alpha}_\gamma\).

{\color{black} Mathematically, specifying a COBS is just picking an orthogonal coordinate system for a vector space equipped with an inner product defined by Eq. \eqref{eq:d1}, with \emph{noncommutative} unit vectors. The exact number of elements of a COBS equals the \emph{square} of the dimensions of the Hilbert space. For example, for a spinless fermion, we can use \(\mathcal{U} = \{\iden, \hat{c}, \hat{c}^\dagger, \hat{\eta} = \hat{c}^\dagger \hat{c} -1/2 \}\); for a spin-1/2, we may choose \(\mathcal{U} = \{\iden,  \si_x, \si_y, \si_z\}\).

For a large class of strongly interacting systems, the model Hamiltonians are defined on a lattice with a finite flavor index (like spin, orbit, etc.). For every lattice point \(\x_{i}\) and flavor there is a local Hilbert space \(\mathscr{H}_{\tau, i}\), where \(\tau\) tracks all flavors. We can denote the local and single flavor COBS as \(\mathscr{U}_{\tau, i} = \{ \hu^{\alpha}_{\tau, i} \}\). Just as the many-body Hilbert space \(\mathscr{H}\) can be constructed as direct product \(\otimes_{\tau,i} \mathscr{H}_{\tau, i}\), the many-body COBS can also be constructed out of the local COBS \(\otimes \mathscr{U}_{\tau, i}\).
 }

With the orthogonality and completeness, any operators, most notably the density operator \(\hrho\), can be expanded in terms of COBS as \(\hrho = \sum_{\alpha} \ev{\hu^\alpha} \hu^{\alpha}\). When we consider equations-of-motion for dynamical correlation functions, i.e. HEOM and SDEOM, they also ensure the \emph{closure of SDEOM}, in principle.

However, for the many body problem, a COBS is exponentially large. We do not expect to solve the problem exactly. We expect to adopt a quantum statistical point of view\cite{Bogolubov-book-intro-quant}, i.e. only a tractable number of correlations, or a small subset of \(\mathcal{U}\), is needed for describing the system, both statically and dynamically.
\subsubsection{COBS for HM: atomic limit}
\label{sec:orgf053f41}
For a single Hubbard atom, we need the following COBS to describe an atomic limit state: i) the single particle operators \(\hat{c}_{\si i}, \hat{c}^\dagger_{\si i}, \heta_{\si i} = \hn_{\si i} - 1/2\), ii) the electronic spin operators \(\hs_{i}^{\alpha} = \hc^\dagger_{i a} \si^{\alpha}_{ab} \hc_{i b}/2\), iii) the local vertex operators \(\hv^\dagger_{\si i} = \hc^\dagger_{\si i}\heta_{\sib i}, ~\hv_{\si i} = \hc_{\si i} \heta_{\sib i}\), iv) the double fermion creation/annihilation operators \(\hc^\dagger_{i \ua}\hc^\dagger_{i \da}, ~\hc_{i \ua}\hc_{i \da}\) and v) the double occupancy operator \(\hat{D}_i = \heta_{\ua i} \heta_{\da i}\). {\color{black} We can verify that the number of COBS elements is 16, corresponding the dimension of the Hilbert space of \(dim = 4\). Note that \(\heta_{\si i}\) double counts with \(\hs^z_i\), but we also need to count the identity operator. }

In the atomic limit, for a half-filled state, we only need to consider \(\ev{\hs^z_i}_0 = \pm 1/2\) as the atomic building blocks. Throughout this work, we use the subscript \(_0\) to denote atomic limit values. Other spin orientations can be rotated to \(\hat{z}\)-direction due to the \(SU(2)\) symmetry. Also, at half-filling \(\ev{\heta_{\ua i}}_0 = -\ev{\heta_{\da i}}_0 = \ev{\hs_i^z}_0\). Therefore, we only need to consider a product state \(\ket{\Psi} = \otimes_i \ket{\ev{\hs^z_i}_0}\) with \(\ev{\hs^z_i}_0 = \pm 1/2\). Note that in the atomic limit, we cannot have a paramagnetic pure state.

When nonlocal correlations are involved, we shall expand the COBS as needed. Then in our notation, the HM at exact half filling is defined by the Hamiltonian
\begin{align}
\mH = \mH_U + \mH_t,\label{eq:d3}
\end{align}
where \(\mH_U = U \sum_i \hat{D}_i\) and \(\mH_t = - \sum_{ij} t_{ij \si} \hc^\dagger_{i \si} \hc_{j \si}\).
\subsection{From HEOM to SDEOM}
\label{sec:org46cf315}

To study the system dynamically, we introduce the usual time-ordered dynamical correlation functions, or simply GFs, between elements of COBS: \(i G_{\pm}[\hu^\alpha(t_i), \hu^\beta(t_f)] = \dev{\mT_\pm \Big( \hu^{\alpha} (t_i), \hu^{\beta} (t_f) \Big)}\), where \(\mT_\pm\) denotes time-ordering with the \(\pm\) sign, \(\dev{~}\) denotes fully dynamical correlations defined as \(\dev{\hu^\alpha (t_i) \hu^\beta (t_f)} = \left\langle (\hu^{\alpha}(t_i)-\ev{\hu^\alpha}) (\hu^\beta(t_f) - \ev{\hu^\beta}) \right\rangle\). For a given state, \(G_{\pm}[\hu^\alpha(t_i), \hu^\beta(t_f)]\) can be solved from the HEOM and SDEOM: \(i \pt_t  G_{\pm}[\hu^\alpha(t_i), \hu^\beta(t_f)] = \ev{[\hu^\alpha, \hu^\beta]_\mp} + \sum_{\gamma \eta} h_\eta b_{\gamma}^{\eta \alpha} \hu^\gamma  G_{\pm}[\hu^\gamma(t_i), \hu^\beta(t_f)]\), in principle. The number of COBS would increase exponentially in practice. Therefore, we need to restrain the calculation to the minimal closed set of SDEOM with controlled approximations. In this work, we shall focus on the single-particle fermionic GFs, \(G_{-}[\hc_{\si_i x_i},  \hc^\dagger_{\si_f x_f}]\), where \(x_i\) and \(x_f\) denote the lattice coordinates of the operator at initial and final times. We also omit the time label unless noted otherwise. For the \(\text{N\'{e}el}\) states at half-filling, we were able to obtain a closed SDEOM with a controlled approximation.
\subsection{Algebraic generation of perturbation sequences.}
\label{sec:org3d266f0}

Traditionally, dynamical perturbation expansion replies on the validity of Wick's theorem thus only applies to canonical operators with a Gaussian ``free'' state. Systematic generation of perturbation sequences in a strong coupling limit, such as for the HM\cite{pairault-2000-stron-coupl,Pairault-1998-Strong-Coupling}, remains a challenging problem.
Within the Hamiltonian and SDEOM approach of this work, the dynamical perturbation sequence on an arbitrary ``free'' state can be generated algebraically through higher orders of time derivatives for the operators and GFs, following Schwinger's construction of action principle \cite{schwinger-1960-unitar-trans}. For the current problem, we need the second order HEOM
\begin{align}
  \begin{split}
    \frac{\pt^2}{\pt t^2} \hO = - \left( [[\mH_t, \mH_U],\hO] + [\mH_t, [\mH_t, \hO]] \right. \\
    \left. + [\mH_t , [\mH_U, \hO] ] +  [\mH_U, [\mH_U, \hO]]\right),
  \end{split}
  \label{eq:d4}
\end{align}
where \(\hO\) is an arbitrary operator of interest.
Eq. \eqref{eq:d4} is exact. When Eq. \eqref{eq:d4} is applied to GFs of \(\mH_U\), it generates the second order dynamical perturbation due to \(\mH_t\). Let \(\hO = \hc_{\si_i i}\), we obtain the dynamical perturbation terms for \(G_{-,0}[\hc_{\si_i x_i},  \hc^\dagger_{\si_f x_f}]\).
\section{ADT solution of atomic Green's Functions}
\label{sec:orgffd03d5}
To obtain a set of solvable equations for \(G_{-,0}[\hc_{\si_i x_i},  \hc^\dagger_{\si_f x_f}]\), the key step is to identify the spectral generating algebra (SGA)\cite{barut-1965-dynam-group} for \(H_{U,i} = U \hat{D}_i\). We find that
\begin{align}
  \begin{split}
    [\hD_i, \hc_{\si i}] = - \hv_{\si i}, \quad [\hD_i, \hc^\dagger_{\si i}] = \hv^\dagger_{\si i},\\
    [\hD_i, \hv_{\si i}] = - \hc_{\si i}/4, \quad [\hD_i, \hv^\dagger_{\si i}] = \hc^\dagger_{\si i}/4,
  \end{split}\label{eq:d5}
\end{align}
which form the minimal set of a closed SGA.
This ensures the HEOM, consequently the SDEOM, are closed.
Accordingly, we take the time derivative twice. Given an atomic state, we find
\begin{align}
  \begin{split}
    & - \pt_t^2 G_{-,0}[\hc_{\si_i x_i},  \hc^\dagger_{\si_f x_f}] =  (\delta'(t) - \si_i \delta(t) U \ev{\hs^z_{i}}_0) \\
    & \times \delta_{\si_i \si_f} \delta_{x_i x_f} + \frac{U^2}{4} G_{-,0}[\hc_{\si_i x_i},  \hc^\dagger_{\si_f x_f}],
  \end{split}\label{eq:d6}
\end{align}
where we used \(\ev{\{\hc_{\si_i x_i},  \hc^\dagger_{\si_f x_f}\}}_0 = \delta_{\si_i \si_f} \delta_{x_i, x_f}\) and \(\ev{\{\hv_{\si_i x_i},  \hc^\dagger_{\si_f x_f} \}}_0 = \ev{\heta_{\sib_i i}}_0 = - \si_i \ev{\hs^z_i}_0\) with a sign convention \(\si = +\) when \(\si = \ua\) and \(\si = -\) when \(\si = \da\).
We simplify the results in frequency space as
\begin{align}
G_{-,0}[\hc_{\si_i x_i},  \hc^\dagger_{\si_f x_f}] = \delta_{x_ix_f} \delta_{\si_i \si_f} \frac{\omega - \si_i U \ev{\hs^z_{x_i}}_0}{\omega^2 - U^2/4}.\label{eq:d7}
\end{align}

Before we proceed to the next part, we point out an important character of the atomic GFs. The \(\ket{\ua}\) state can only be excited by \(\hc_{\ua}\) or \(\hc^\dagger_{\da}\), not the other way around. In terms of the GFs, we find that \(\ev{\hs^z_i}_0 = 1/2\), we find \(G_{-,0}[\hc_{\si_i x_i},  \hc^\dagger_{\si_f x_f}] = (\omega + U/2)^{-1}\) and for \(\ev{\hs^z_i}_0 = -1/2\) \(G_{-,0}[\hc_{\si_i x_i},  \hc^\dagger_{\si_f x_f}] = (\omega - U/2)^{-1}\). Therefore, there is only one pole with nonzero spectral weight in the atomic limit, which corresponds to either the upper or lower Hubbard band (UHB or LHB). Such behavior also can be viewed as a Luttinger ``surface'' or GF zeros\cite{dzyaloshinskii-2003-some-conseq,dave-2013-absen-luttin} in the atomic limit. We shall find this character still manifesto in the spin-projected GFs of the lattice case.
\section{Magnetic energy in the large-U limit}
\label{sec:orga2285d1}

Next, we employ Eq. \eqref{eq:d5} to prove within the current framework that the spins should be antiferromagnetically aligned. To achieve that, we compute the expectation value of the nn hopping term \(\ev{\hc^\dagger_{\si i} \hc_{\si i+1}} = - \ev{\hc_{\si i+1} \hc^\dagger_{\si i}}\) perturbatively to the second order. We compute via the relation \(\ev{\hc_{\si i} \hc^\dagger_{\si j}} = iG_{-}[\hc_{\si i}, \hc^\dagger_{\si j}](t = 0^+)\).

By inspecting the operator algebras, we find the lowest order nonzero correction to be \(-\pt^2_t G_{-}[\hc_{\si i}, \hc^\dagger_{\si j}] = - t U \ev{\heta^z_{\sib i}} G_{-,0}[\hc_{\si j} , \hc^\dagger_{\si j}].\) The correction comes from \([H_t,[H_U, \hc_{\si i}]]\) of Eq. \eqref{eq:d5}.
Plug in the results of \eqref{eq:d3}, we find \(\ev{\hc_{\si i} \hc^\dagger_{\si j}} = -\frac{t}{U} \ev{\heta^z_{\sib i}}_0 \ev{\heta^z_{\si j}}_0\). Again, use \(\ev{\heta^z_{\si i}} = - \ev{\heta^z_{\sib i}} = \ev{\hs^z_i}\) and the correction to energy per bond is \(\Delta E = \frac{4 t^2 \ev{\hs^z_i}_0 \ev{\hs^z_j}_0 }{U}\). The sign shows that \(\ev{\hs^z_i}_0\) aligned in an AFM pattern is energetically more favorable when subject to the perturbation of \(\mH_t\). The amplitude \(4t^2/U\) is consistent with previous results through the superexchange mechanism\cite{anderson-1950-antif,ding-2019-effec-exchan}. For previous works in the strong coupling limit\cite{kotliar-1986-new-funct,yoshioka-1989-slave-fermion,han-2019-finit-temper,lee-2006-dopin-mott-insul,florens-2002-quant-impur}, capturing magnetism consistently is a unmet challenge until the current work.

\textcolor{black}{
With this magnetic energy, we can directly lift the \(2^V\)-fold degeneracy of the atomic limit ground state, down to a two-fold degeneracy of \(\text{N\'{e}el}\) states on a bipartite lattice. However, assuming a \(\text{N\'{e}el}\) state automatically implies that we have to take the thermodynamic limit now so that a spontaneous symmetry breaking is allowed. Taking the thermodynamic limit also allows us to work in the momentum space in later sections.
}
\section{Second order theory from \(\text{N\'{E}El}\) states.}
\label{sec:org0478cd2}

In this part, we solve for \(G_{-}[\hc_{\si_i x_i},  \hc^\dagger_{\si_f x_f}]\) at the second order on top a product AFM \(\text{N\'{e}el}\) state. However, we must emphasize that the AFM \(\text{N\'{e}el}\) state is still two-fold degenerate. A more rigorous calculation requires parameterization of the two-fold degeneracy before the perturbation. We restrict to the product state for a simpler demonstration.

Consider the full SDEOM up to the second order time-derivative:
\begin{align}
  \begin{split}
    & \omega^2 G_{-}[\hc_{\si_i i},\hc^\dagger_{\si_f f}]  = \omega \delta_{x_i,x_f} \delta_{\si_i \si_f}\\
    &  + U \omega G_{-}[\hv_{\si_i x_i}, \hc^\dagger_{\si_f x_f}]  - \sum_j t_{ij} \omega G_{-}[\hc_{\si_i j},\hc^\dagger_{\si_f f}]
  \end{split}\label{eq:d8}
  \\
  \begin{split}
    & \omega G_{-}[\hv_{\si i},\hc^\dagger_{\si f}]  =  -\si\ev{\hs^z_i}_0 \delta_{x_i,x_f} \delta_{\si_i \si_f}  \\
    & + U/4 G_{-}[\hc_{\si i},\hc^\dagger_{\si_f f}]  - \sum_j t_{ij} (G_{-}[\heta_{\sib_i i} \hc_{\si_i j},\hc^\dagger_{\si_f f}] \\
    &+ G_{-}[\hs^+_i\hc_{\sib_i j},\hc^\dagger_{\si_f f}]  - G_{-}[\hc^\dagger_{\sib_i j} \hc_{\si_i i} \hc_{\sib_i i}, \hc^\dagger_{\si_f f}] ),
  \end{split}\label{eq:d9}
\end{align}
where the unperturbed values are used for static correlations.

To obtain closed equations, we make further approximations on the new GFs generated in Eq. \eqref{eq:d7}. First, we argue that we can simply ignore both \(G_{-}[\hs^+_i\hc_{\sib_i j},\hc^\dagger_{\si_f f}] \text{ and } G_{-}[\hc^\dagger_{\sib_i j} \hc_{\si_i i} \hc_{\sib_i i}, \hc^\dagger_{\si_f f}]\). \(G_{-}[\hs^+_i\hc_{\sib_i j},\hc^\dagger_{\si_f f}]\) is the coupling between single electron excitations and AFM spin fluctuations, which is apparently a higher order effect. The lowest order and dominating effect should be the AFM order, not its fluctuations. \(G_{-}[\hc^\dagger_{\sib_i j} \hc_{\si_i i} \hc_{\sib_i i} , \hc^\dagger_{\si_f f}]\) can be estimated as \(\ev{\hc^\dagger_{\sib_i j}\hc_{\sib_i i}} G_{-}[ \hc_{\si_i i}, \hc^\dagger_{\si_f f}]\) near the atomic limit, which is a self-iteration of the single-particle GFs hence also a higher order term.

Therefore, we are left with the non-local vertex function \(G_{-}[\heta_{\sib_i i} \hc_{\si_i j},\hc^\dagger_{\si_f f}]\) which we approximate in a Hartree\cite{hartree-1928-wave-mechan} way
\begin{align}
G_{-}[\heta_{\sib_i i} \hc_{\si_i j},\hc^\dagger_{\si_f f}] \simeq \ev{\heta_{\sib i}}_0 G_{-}[\hc_{\si_i j},\hc^\dagger_{\si_f f}].\label{eq:d10}
\end{align}
This approximation becomes exact when the unperturbed state is fully spin-polarized hence exact for the \(\text{N\'{e}el}\) product state. To deal with  \(G_{-}[\heta_{\sib_i i} \hc_{\si_i j},\hc^\dagger_{\si_f f}]\) more accurately and generically, we will need to consider the parent state with spatial spin entanglement, i.e. as a partially polarized, superposition of the two degenerate AFM states. We leave that for the future.

To the simplest but reasonable approximations and after eliminating \(G_{-}[\hv_{\si i},\hc^\dagger_{\si f}]\), we have
\begin{align}
  \begin{split}
  & \left(\omega^2 - \frac{U^2}{4} \right) G_{-}[\hc_{\si_i i},\hc^\dagger_{\si_f f}] + \sum_j t_{ij} (\omega - \si_i \ev{\hs^z_i} U ) \\
  & \times G_{-}[\hc_{\si_i j},\hc^\dagger_{\si_f f}] = (\omega  -\si \ev{\hs^z_i} U) \delta_{x_i,x_f} \delta_{\si_i \si_f},
  \end{split}\label{eq:d11}
\end{align}
where we used \(\ev{\heta^z_{\sib_i i}} = - \si_i \ev{\hs^z_i}\). Note that from here on, we drop the \( _0 \) index and promote \(\ev{\hs^z_i}\) to be a self-consistent state parameter.

Finally, a spatial-momentum Fourier transform (FT) makes the equation invertible by expressing \(\ev{\hs^z_i} = m_z e^{i \x_i \cdot \vQ}\) with \(\vQ = (\pi, \pi)\):
\begin{align}
  \begin{split}
   &  (\omega^2 - U^2/4 - \omega \epsilon_{\k_i} )) G_{-}[\hc_{\si_i \k_i},\hc^\dagger_{\si_f \k_f}] \\
   & - U m_z \epsilon_{\k_i - \vQ} G_{-}[\hc_{\si_i \k_i - \vQ},\hc^\dagger_{\si_f \k_f}] \\
   & = (\omega - \si_i m_z U) \delta_{\k_i, \k_f} \delta_{\si_i \si_f}.
  \end{split}\label{eq:d12}
\end{align}

Eliminating \(\k_f\) and keeping only diagonal spin indices (letting \(\sigma_i = \sigma_f = \sigma\)), we find the \(\vQ\) zone-folding equations for \(G_{-}[\hc_{\si \k},\hc^\dagger_{\si \k}]\). Let \(G_{-}[\hc_{\si \k},\hc^\dagger_{\si \k}] = G_{kk},  G_{-}[\hc_{\si \k - \vQ},\hc^\dagger_{\si \k}] = G_{Qk},  G_{-}[\hc_{\si \k - \vQ},\hc^\dagger_{\si \k - \vQ}] = G_{QQ}\). We also use the square lattice nn bare dispersion \(\epsilon_{\k} = -2 t (\cos k_x + \cos k_y)\) which satisfies \(\epsilon_{\k+\vQ} = - \epsilon_{\k}\).
{\color{black} This simplification will not hold if other hoppings, such as the next-nearest-neighbor term, are considered.}
Note that the sign-changing behavior would be spoiled if other hopping terms are included. Finally, we have
\begin{align}
  \begin{split}
    & \begin{pmatrix}
      \omega^2 - U^2/4 - \omega \epsilon_{\k} & - U m_z \epsilon_{\k} \\
      U m_z \epsilon_{\k} & \omega^2 - U^2/4 + \omega \epsilon_{\k}
    \end{pmatrix}\\
    \times
    & \begin{pmatrix}
        G_{kk} & G_{kQ} \\ G_{Qk} & G_{QQ}
      \end{pmatrix} = \begin{pmatrix} \omega - \si m_z U & 0 \\ 0 & \omega - \si m_z U \end{pmatrix}
  \end{split}\label{eq:d13}
\end{align}
which can be solved by matrix inversion. For the diagonal GF, we obtain
\begin{align}
G_{kk} = \frac{(\omega - \si m_z U) \left(\omega^2 -U^2/4 - \epsilon_{\k} \omega \right)}{(\omega^2 -U^2/4)^2 - \epsilon_{\k}^2 (\omega^2 - m_z^2 U^2) },\label{eq:d14}
\end{align}
\begin{align}
G_{kQ} = \frac{(\omega + \si m_z U) \sigma \epsilon_{\k} m_z U}{(\omega^2 -U^2/4)^2 - \epsilon_{\k}^2 (\omega^2 - m_z^2 U^2) }.
\end{align}
We can identify \(G_{kk}\) and \(G_{QQ}\) as GFs within the same sublattice (\(A \rightarrow A\) or \(B \rightarrow B\)), and \(G_{kQ}\) and  \(G_{Qk}\) as the GFs between different sublattices ((\(A \rightarrow B\) or \(B \rightarrow A\) ). \(G_{QQ} (G_{Qk})\) can be obtained from \(G_{kk} (G_{kQ})\) by letting \(\sigma \rightarrow - \sigma\) and \(\epsilon_{\k} \rightarrow - \epsilon_{\k}\).
Note that in the final expression Eq. \eqref{eq:d14} we have restored all \(m_z\) so that we can treat it as a self-consistent equation which would allow us to determine the state parameters. In this case, we only have a single parameter \(m_z\). With these understanding, it is sufficient to study \(G_{kk}\).
\section{Application}
\label{sec:orgcbe8fe8}
With Eq. \eqref{eq:d14}, we study and discuss various major aspects of AFM Mott insulators which observed both experimentally and numerically.
\subsection{Self-consistent solution}
\label{sec:orgca5aa4a}

\begin{figure}
  \centering
  \subfigure[]{\label{sfig:1a}\includegraphics[width=0.45\columnwidth]{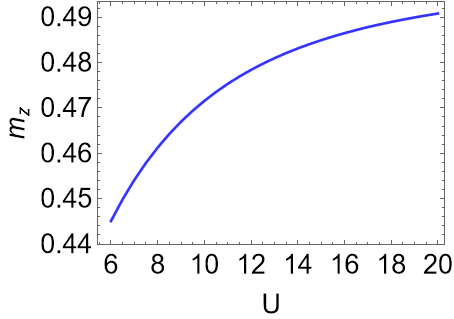}}
  \subfigure[]{\label{sfig:1b}\includegraphics[width=0.45\columnwidth]{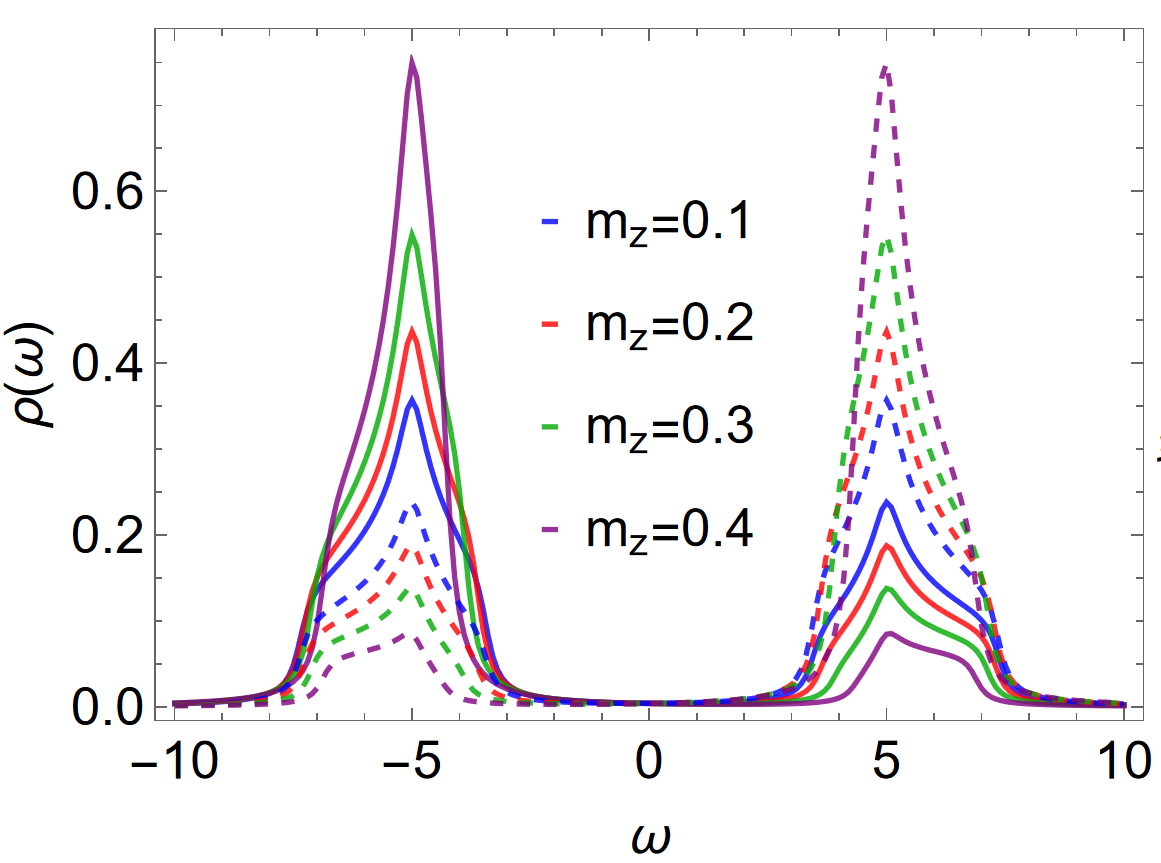}}
  \caption{\ref{sfig:1a} Self-consistent solution to magnetization. Here only single-particle charge fluctuation is taken into account, thus the deviation from full polarization is small. \ref{sfig:1b} Spin projected local density of states at different magnetization. The solid line is for spin up and the dashed line is for spin down. To exemplify the blueshift effect and the spectral imbalance, we show results for \(m_z\) down to 0.1 beyond the self-consistenct results.}
\end{figure}

With the lattice GF Eq. \eqref{eq:d12}, we can compute \(m_z\) according to \(\ev{\hs^z_i} = \ev{\eta_{\si i}} = \ev{[\hc_{\si \i}, \hc^\dagger_{\si \i}]}/2 = (1/2) i G_{-}[\hc_{\si \i}, \hc^\dagger_{\si \i}](t=0) = - (1/2) \int d^2k (2\pi)^{-2} \Im m[G_{-}[\hc_{\si \k}, \hc^\dagger_{\si \k}]\). Without actually carrying out the \(\k\)-integral, we can expand Eq. \eqref{eq:d12} first and check the sum rule. We find a self-consistent equation \(m_z = \int d^2k (2\pi)^{-2} (m_z - m_z/U + f(m_z^2))\). So, to the linear order, we do find that \(m_z\) decreases with decreasing \(U\) and the higher order terms allow us to obtain a self-consistent solution for \(m_z\). We numerically solve the self-consistent equation and plot \(m_z\) in Fig. \ref{sfig:1a}. Therefore, \(m_z\) only reaches the result of AFM Heisenberg model value\cite{sandvik-1997-finit-size} in the \(U \rightarrow \infty\) limit. At finite \(U\), we obtain an additional \(1/U\) correction on top of that.

We plot the spin projected local density of states (LDOS), partly also to illustrate the Mott gap blueshift, which agree well with cluster dynamical mean field results in Ref. \cite{fratino-2017-signat-mott}. Here we manually set the magnetization all the way down to \(m_z = 0.1\) to exemplify the blueshift effect and the spectral imbalance, although such low magnetization regime is not accessible from the self-consistent solution.
\subsection{Mott Gap}
\label{sec:org6dfda7d}
\begin{figure}
  \centering
  \includegraphics[width=0.85\columnwidth]{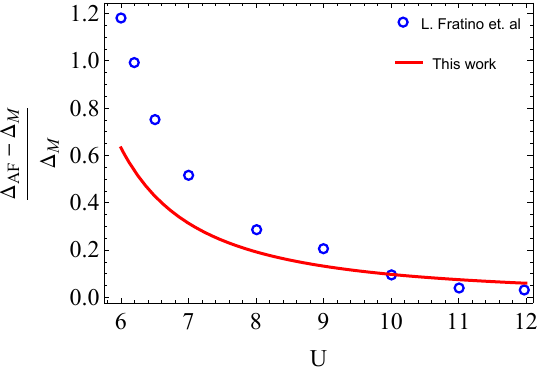}
  \caption{\ref{sfig:2} A direct comparison of \(\Delta_{bs}/\Delta_{Param}\) of this work (red line) and data extracted from Fig.(3d) of Ref. \cite{fratino-2017-signat-mott}(blue dots). We modified \(m_z(U)\) by a constant factor as \(\rightarrow m_z(U)*1.73/1.94\) to match the staggered magnetization of Ref. \cite{fratino-2017-signat-mott} in the large-U limit since our computed \(m_z(U)\) lacks both spin-wave and thermal fluctuations.}
  \label{sfig:2}
\end{figure}

Now we turn to the spectral properties. First of all, we obtain four dispersive upper or lower Hubbard bands from the poles of Eq. \eqref{eq:d14}:
\begin{align}
  \begin{split}
    \omega_{0,i} &= 
              \pm \frac{U}{2} \sqrt{1 + \tilde{\epsilon}_{\k}^2 \pm \abs{\tilde{\epsilon}_{\k}} \sqrt{\tilde{\epsilon}_{\k}^2 + 2(1 - 4 m_z^2)}},
  \end{split}\label{eq:d13}
\end{align}
where \(\tilde{\epsilon}_{\k} = \sqrt{2} \epsilon_{\k} / U\) is a dimensionless, renormalized bare band dispersion. To extract the direct Mott gap, we can expand the dispersion Eq. \eqref{eq:d13} in the large-\(U\) limit, which yields \(\omega_{0,i} \simeq \pm (U/2 - \sqrt{1 - 4 m_z^2} \epsilon_{\k}/2)\). The band gap is minimal at \(\epsilon_{\k} = W\) where \(W\) is the half bandwidth. For HM on the 2D square lattice with only nn hopping, \(W = 4t\). Therefore, we find the Mott gap is \(\Delta_{\text{Mott}} = U - \sqrt{1 - 4 m_z^2}~W\). The gap size increases with the AFM moment \(m_z\). Such antiferromagnetism induced gap increasing is known as the magnetic blueshift for Mott insulators\cite{blazey-1969-antif-phase,wang-2009-antif-gap,fratino-2017-signat-mott,hafez-torbati-2021-magnet-blue,hafez-torbati-2022-simpl-approac}, which were observed both experimentally and numerically. Our results give the blueshift of the Mott gap due to AFM order to be \(\Delta_{bs} = \Delta_{\text{AFM}} - \Delta_{\text{Param}} = (1-\sqrt{1 - 4 m_z^2})~W\). We immediately find it indeed increases \(m_z\), in agreement with both numerical calculations and experimental observations.
{\color{black}
A direct comparison between \(\Delta_{bs}/\Delta_{Param}\) of this work and data extracted from Fig.(3d) of Ref. \cite{fratino-2017-signat-mott} is shown in Fig. \ref{sfig:2}. We modified \(m_z(U)\) by a constant factor as \(\rightarrow m_z(U)*1.73/1.94\) to match the staggered magnetization of Ref. \cite{fratino-2017-signat-mott} in the large-U limit since our computed \(m_z(U)\) lacks both spin-wave and thermal fluctuations.
}
\subsection{Dispersion and spectral weights}
\label{sec:org5cffd43}

To carefully examine the properties of the Hubbard bands, we rewrite Eq. \eqref{eq:d14} in terms of the poles \(\omega_{0,i}(\k)\) and their corresponding weights \(\mathcal{W}_{i}(\k)\):
\begin{align}
  \begin{split}
    & G_{-}[\hc_{\si \k}, \hc^\dagger_{\si \k}](\omega, \k) = \sum_i \frac{\mathcal{W}_{i}(\k)}{\omega - \omega_{0,i}(\k)} \\
    & = \sum_i \frac{A_{i}(\k)}{\omega - \omega_{0,i}(\k) + i 0^+} + \frac{B_{i}(\k)}{\omega - \omega_{0,i}(\k) - i 0^+},
  \end{split}\label{eq:d17}
\end{align}
where \(i = 1, \dots, 4\) indicates the four Hubbard bands. While a large-\(U\) splits the non-interacting band into two, the presence of AFM order doubles that by Brillouin-zone-folding.
We find that \(\mathcal{W}_{i}(\k)\) is not a positive-definite function. This is because \(G_{-}[\hc_{\si \k}, \hc^\dagger_{\si \k}](\omega, \k)\) is a time-ordered GF, containing both retarded and advanced parts. \(\sgn(\mathcal{W}_i)\) automatically distinguishes between the retarded and advanced parts. Therefore, in the second line of Eq. \eqref{eq:d14}, we choose \(A_i = \Theta(\mathcal{W}_i) \mathcal{W}_i, ~ B_i = \Theta(-\mathcal{W}_i) (-\mathcal{W}_i)\), in accordance with conventional notations so that both \(A_i\) and \(B_i\) are positive-definite functions. Thus the spectral function of \(G_{kk}[\hc_{\si \k}, \hc^\dagger_{\si \k}]\), which is denoted as \(\mathcal{A}_{kk \sigma}(\omega, \k) = - \pi^{-1} \Imm[G_{kk}]\) is given as: \(\mathcal{A}_{kk}(\omega, \k) \sim i(G^R - G^A) = \sum_i \delta(\omega - \omega_{0,i}(\k)) (A_{i}(\k) + B_{i}(\k))\).

\begin{figure*}
    \centering
    \subfigure[]{
      \includegraphics[width=0.95\columnwidth]{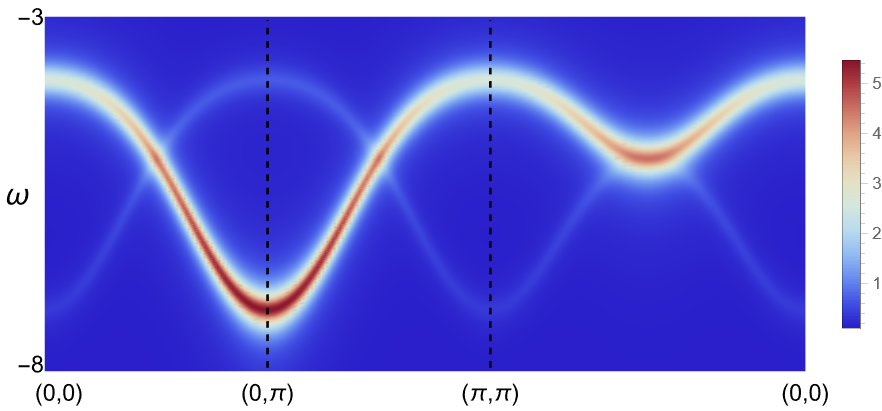}
      \label{sfig:2a}
    }
    \subfigure[]{\label{sfig:2b}\includegraphics[width=0.95\columnwidth]{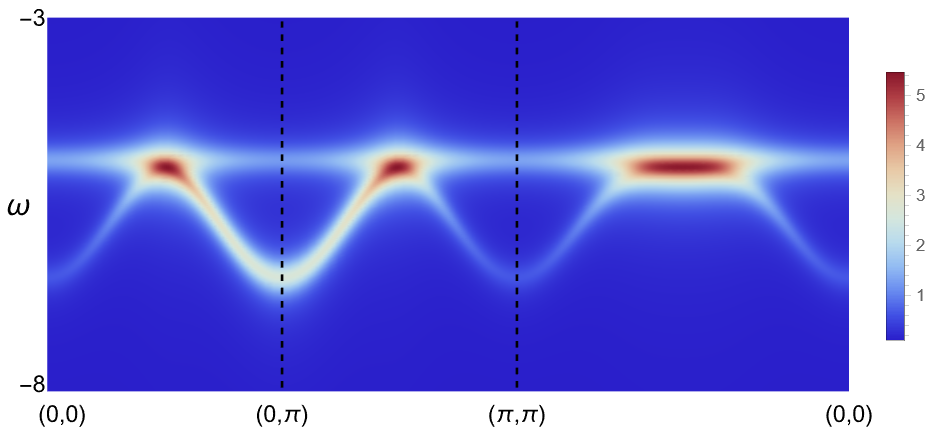}}
    \subfigure[]{\label{sfig:2c}\includegraphics[width=1.95\columnwidth]{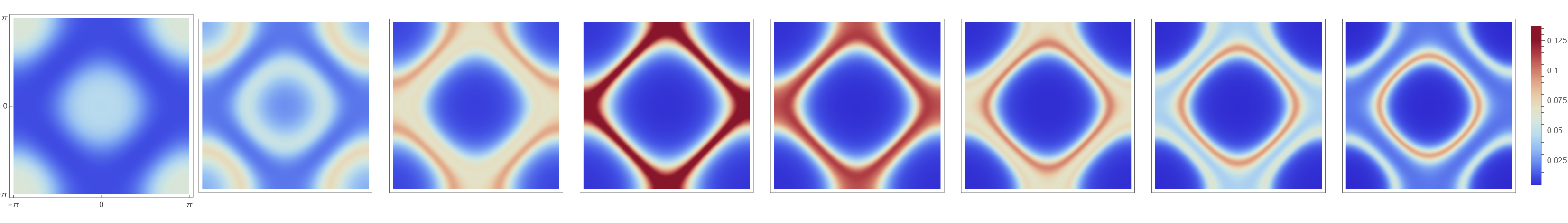}}
    \subfigure[]{\label{sfig:2d}\includegraphics[width=1.9\columnwidth]{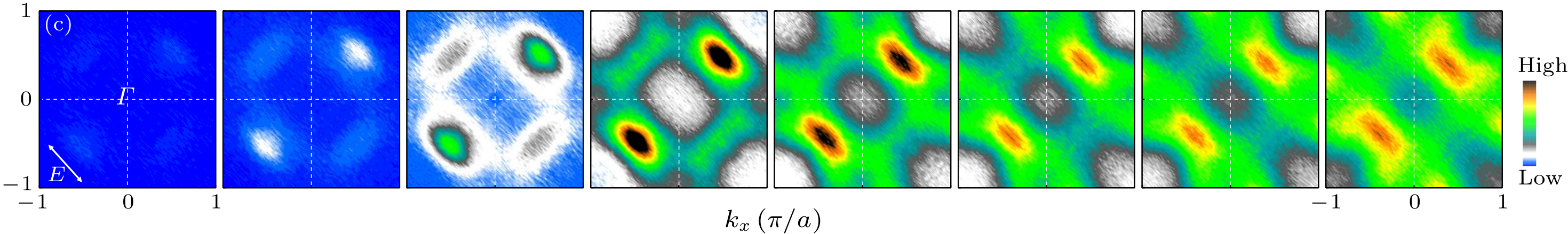}}
    \caption{\ref{sfig:2a} MDC for \(m_z = 0.45\) of \(\mathcal{A}[\hat{c}_{\k \uparrow}, \hat{c}^\dagger_{\k \uparrow}]\), i.e. that of a single spin species. \ref{sfig:2b} MDC at \(m_z = 0.49\). Both are plotted along high symmetry directions \((0,0)\rightarrow (0,\pi) \rightarrow (\pi,\pi) \rightarrow (0,0)\). \ref{sfig:2c} Constant energy cuts at \(\omega = - U/2 + \delta\omega\) with \(\delta\omega \in \{1.5t, t, 0.5t, 0.2t, 0.1t, -0.2t\} \), from left to right, for \(m_z = 0.45\). \ref{sfig:2d} Constant energy cuts from recent APRES measurements (Fig. (1c) of Ref. \cite{hu-2018-eviden-multip}) on a cuprate parent compound.
    \ref{sfig:2b} showcase the interesting limit where the fully polarized local moments turn one of the LHB into an exact flat band. \ref{sfig:2c} is comparable to ARPES observations of Ref. \cite{hu-2018-eviden-multip}. Plot for \(\delta \omega = -U/2 + 0.2 t\) provides an explanation for the two "Fermi surface sheets" observed.}\label{fig:d2}
\end{figure*}

\begin{figure*}
\centering
 \subfigure[VB cuts]{\label{sfig:4a}\includegraphics[width=1.9\columnwidth]{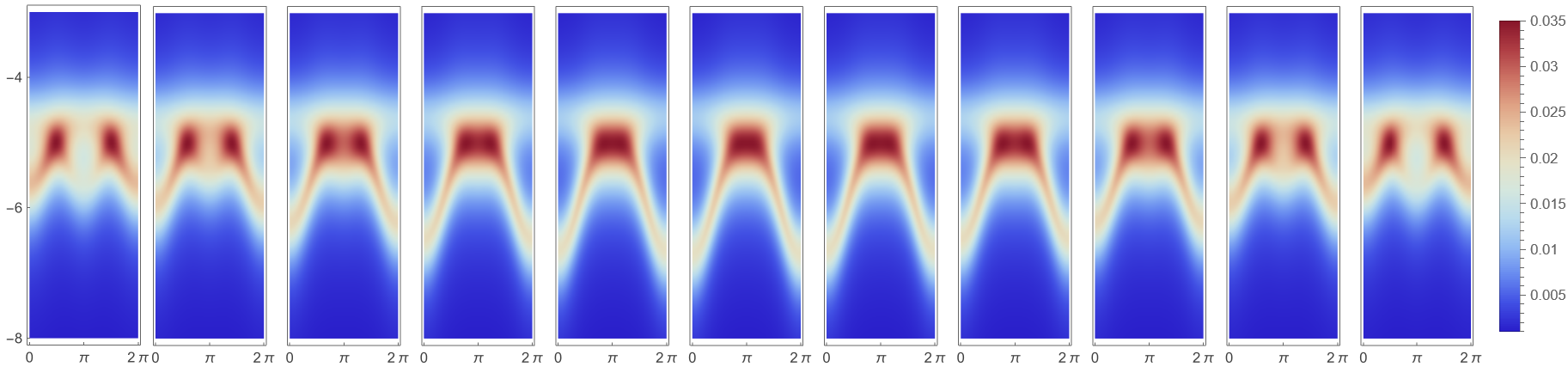}}
 \subfigure[HB cuts]{\label{sfig:4b}\includegraphics[width=1.9\columnwidth]{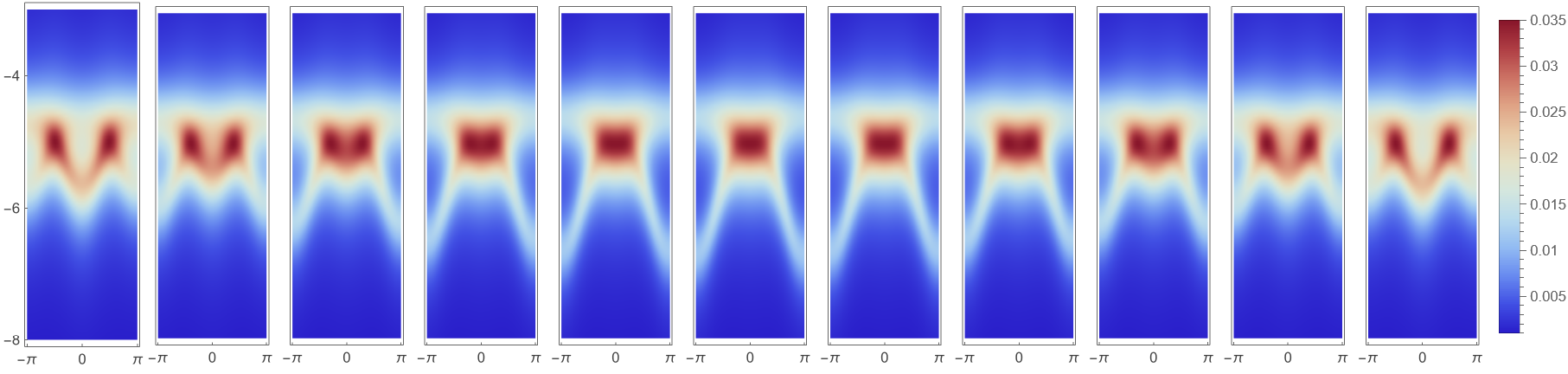}}
 \subfigure[]{\label{sfig:4c}\includegraphics[width=1.9\columnwidth]{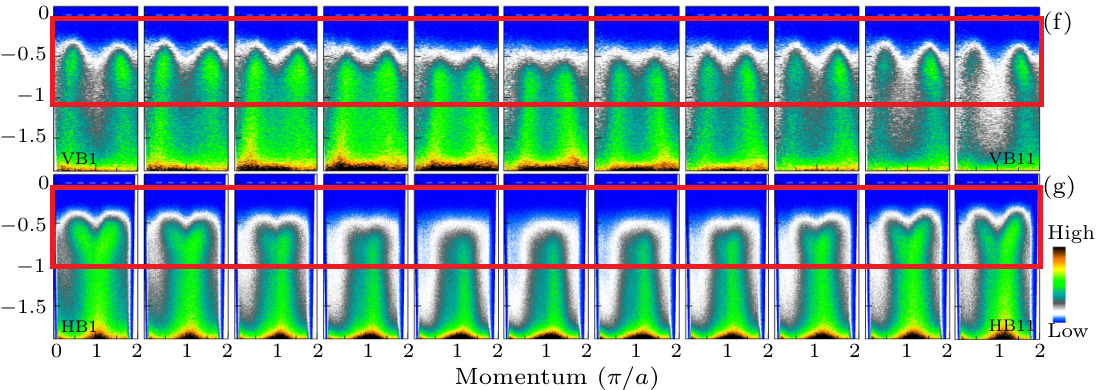}}
 \caption{Comparison of MDCs between this work and Ref. \cite{hu-2018-eviden-multip}. \ref{sfig:4a} MDCs along the cuts VB1 to VB11 and \ref{sfig:4b} HB1 to HB11. \ref{sfig:4c} ARPES data of MDC along the same cuts. Cuts are defined the same as in Ref. \cite{hu-2018-eviden-multip}. Cuts labeled as VB1 to VB11 are given as \(\{dk_x, k_y\}\) with \(k_y \in (0, 2 \pi)\) and \(dk_x \in \{-\pi/2,-\pi/2 + \pi/10, \dots, \pi/2\}\). The HB cuts are given as \(\{k_x, dk_y\}\) with \(k_x \in (-\pi, \pi)\) and \(dk_y \in \{\pi/2, \pi/2 + \pi/10, \dots, 3\pi/2\}\). To make a direct comparison, we note that the current results of LHB have a renormalized bandwidth \(\sqrt{1-4 m_z^2} 4t \simeq 0.6 \) eV, assuming realistic parameters with \(t \simeq 0.4~eV\)\cite{lebert-2023-param-disper} and \(m_z \simeq 0.3\)(60\% polarization) for the 2D square lattice Heisenberg model\cite{sandvik-1997-finit-size}. Therefore, our results are directly comparable to the ARPES measurement within an energy window of \(0.6~eV\), which is marked by red frames in \ref{sfig:4c}.}\label{sfig:4}
 \end{figure*}

To compare with ARPES-measured spectral functions directly, we consider the total spin-averaged spectral function \(\mathcal{A}(\omega, \k) = \sum_{k_1,k_2, \sigma} \mathcal{A}_{k_1 k_2 \sigma}\) with \(k_{1(2)} = k,~Q\) and \(\sigma = \ua,~\da\). We plot \(\mathcal{A}(\omega, \k)\) numerically in both Fig. \ref{fig:d2} and \ref{sfig:4}
{\color{black}
with Lorentzian factors \(\delta = 0.3\) and \(\delta = 0.5\) respectively.
Although the theoretical results are obtained at zero temperature, we expect they can still well describe the system at low temperatures. The reasons are as follows: For the fully gapped system we considered, the energy scale relevant to the lowtemperature properties is at the order of \(U\) or \(J\); for cuprates, either ( \(U, J>0.1 \mathrm{eV}\) ) is at least one order of magnitude larger than the temperature at which the measurements were taken ( \(T \lesssim 100 \mathrm{~K}\) ). Therefore, we expect the low-temperature spectral features to be similar to the zero-temperature ones up to a Lorentzian broadening factor.
}

In Fig. \ref{fig:d2}, we first show a generic momentum distribution curve (MDC) along high symmetry directions \((0,0)\rightarrow (0,\pi) \rightarrow (\pi,\pi) \rightarrow (0,0)\) for \(m_z = 0.45\) in \ref{sfig:2a}, then the extreme case for \(m_z = 0.49\) in \ref{sfig:2b}. \ref{sfig:2b} showcase the interesting limit where the fully polarized local moments turn one of the LHB into an exact flat band. We plot constant energy cuts at \(\omega = - U/2 + \delta\omega\) with \(\delta\omega \in \{1.5t, t, 0.5t, 0.2t, 0.1t, -0.2t\}\), from left to right, at \(m_z = 0.45\). In all plots we set \(U =10t\) and \(t=1\). Recent ARPES studies on the parental compounds revealed nontrivial dispersion of the lower Hubbard bands\cite{hu-2018-eviden-multip}. Our results at \(\delta \omega = -U/2 + 0.2 t\) provides a direct explanation for the ``two Fermi surface sheets'' observed.

In Fig. \ref{sfig:4}, we plot MDCs in two series of cuts in \(\omega-\k\) planes. Cuts are defined the same as in Ref. \cite{hu-2018-eviden-multip}. Cuts labeled as VB1 to VB11 are given as \(\{dk_x, k_y\}\) with \(k_y \in (0, 2 \pi)\) and \(dk_x \in \{-\pi/2,-\pi/2 + \pi/10, \dots, \pi/2\}\). The VH cuts are given as \(\{k_x, dk_y\}\) with \(k_x \in (-\pi, \pi)\) and \(dk_y \in \{\pi/2, \pi/2 + \pi/10, \dots, 3\pi/2\}\).

To make a direct comparison, we note that the current results of LHB have a renormalized bandwidth \(\sqrt{1-4 m_z^2} 4t \simeq 0.6~eV\) , assuming realistic parameters with \(t \simeq 0.4~eV\)\cite{lebert-2023-param-disper} and \(m_z \simeq 0.3\)(60\% polarization) for the 2D square lattice Heisenberg model\cite{sandvik-1997-finit-size}. Therefore, our results are directly comparable to the ARPES measurement within an energy window of \(0.6~eV\), which is marked by red frames in Fig. \ref{sfig:4c}. The spectral
{\color{black}
beyond
}
this energy window\cite{ronning-2005-anomal-high} may be attributed to 3-band physics, such as the Zhang-Rice singlets\cite{zhang-1988-effec-hamil}.

Once the appropriate energy window for comparison is identified, one immediately recognize that the prominent features of theoretical spectral functions from \eqref{eq:d14} match the APRES results well.
For the VB cuts, the high intensity regimes near \(dk_x = \pm \pi/2\) form a split ``\(\wedge\)'' shape. As \(dk_x \rightarrow 0\), the split narrows. For the HB cuts, the high intensity regimes near \(dk_y = \pi/2\) or \(3\pi/2\) form a ``\(\vee\)'' shape. As \(dk_y \rightarrow \pi\) (from left \& right towards the middle), the split of the ``V''-shape also narrows, forming a seemingly isolated patch. While for the theoretical results, this narrowing behavior seems symmetric between VB and HB cuts, the ARPES results are asymmetric, making the HB cuts comparison better. This difference may be attributed to a potential short-range magnetic anisotropy which is absent in the current calculation.
\subsection{Luttinger points and surfaces}
\label{sec:org31692a8}

Luttinger points and surfaces\cite{dzyaloshinskii-2003-some-conseq} are believed to play an important role in both the undoped Mott insulating phases and more importantly the doped systems\cite{stanescu-2007-theor-luttin,yang-2022-violat-luttin,yang-2006-phenom-theor}.
At half-filling, they are at least relevant to three different aspects. First, such zeros can be re-cast as divergences of self-energies\cite{schaefer-2013-diver-precur}.
Secondly, GF's zeros contribute to Luttinger's theorem on counting the Fermi volumes for interacting systems.\cite{dave-2013-absen-luttin,seki-2017-topol-inter,setty-2024-elect-proper}.
Finally, GF zeros are essential for the topology of Mott insulators \cite{raghu-2008-topol-mott-insul,rachel-2010-topol-insul,mai-2023-is-new,sen-2020-mott-trans}. However, all the above aspects are topological in a certain perspective.
Our work provides an analytic case from a microscopic model, which makes a more concrete discussion of these topological aspects of single band Mott insulators.

We must note that there are two different types or definitions of zeros in GFs: (i) \(G_{\tau, \tau'} = 0\)  and (ii) \(\det[ G_{\tau \tau'} ] = 0\), where \(\tau\)s can be either orbital or/and spin indexes. For the AFM Mott states we obtain in this work, we have two \(\tau\)s: i) the spin index \(\sigma\) and the sublattice index \(\tau = A\) or \(B\). When being used for discussing topological properties, both types of zeros could be used albeit definition (ii) is more commonly used for interacting systems. However, a recent work\cite{misawa-2022-zeros-green} showed that zeros of single (diagonal-in-orbital) component of GFs also can be used to identify the underlying topological phases. We briefly discuss aforementioned topological aspects with the relevant types of zeros but leave detailed studies to future works.

First, we consider the zeros of the diagonal components according to Eq. \eqref{eq:d14}. The Eq. \eqref{eq:d14} clearly possesses two distinct ``bands'' of zeros. The first zero band is \(\omega = \pm m_z U\) which is already present in the atomic limit. Within the current solution, it remains dispersion-less. However, if we inspect its origin, the right-hand side (RHS) of Eq. \eqref{eq:d11} which contains static correlation functions\cite{Ding-2022-algeb-dynam}, we should expect it to acquire some dispersion if we further include higher orders of correlation functions. The second zero band is given by \(\omega^2 -U^2/4 - \epsilon_{\k} \omega = 0\) and is dispersive. To understand its origin, we take a \(m_z \rightarrow 0\) limit. The factor \(\omega^2 -U^2/4 - \epsilon_{\k} \omega\) is immediately canceled by the denominator. Apparently, the second band of zeros is a result of quasi-particle band splitting due to AFM-caused Brillouin zone folding.

To understand the first zero band further, we rewrite Eq. \eqref{eq:d14} in a traditional self-energy form and analyze it as the Eq. (1) of Ref. \cite{wagner-2023-mott-insul} (\(\Sigma( k , \omega)=\frac{U^2 / 4}{\omega+\widetilde{H}_0( k )}\)). By setting \(m_z \rightarrow 0,~ \sigma \rightarrow 1\), we have \(G_{-}[\hc_{\si \k}, \hc^\dagger_{\si \k}] = (\omega + \epsilon_{\k} - U^2/(4\omega) )^{-1} ~  \rightarrow \Sigma( k , \omega) = \frac{U^2 / 4}{\omega}\).
{\color{black}
Therefore, this analysis of our GFs gives \(\widetilde{H}_0( k ) = 0\).
}
This can be expected since the \(m_z \rightarrow 0\) limit ignores all magnetic correlations, which are the lowest order of spatial fluctuations that are allowed in the Mott phase. Hence this Luttinger surface at \(\omega = 0\) is purely a local effect caused by a sufficiently large \(U\) which is seen in DMFT calculations\cite{schaefer-2013-diver-precur} as a diverging self-energy.
{\color{black}
Furthermore, we expect this band of zeros to acquire dispersion if we could incorporate spatial magnetic fluctuations.
}

In contrary, the second band of zeros originates from both the bare band dispersion and the AFM-induced zone folding. Note that it would be completely canceled if we set \(m_z \rightarrow 0\). Therefore, it is more relevant to the Hubbard-band topology. A systematical understanding of the topology Hubbard-bands would require equal-footing treatment of both zeros and poles. Although some expressions of topological invariants, such as Ishikawa-Matsuyama formula\cite{ishikawa-1986-magnet-field} and Volovik's \(N_3\)\cite{book-Volovik-2009-unive-in-helium}, do contain contribution from zeros, they are nonetheless generalized from non-interacting band systems and are restricted to a certain class of topological phases. It is not clear whether these formulas could correctly characterize topology in the presence of both types of zero bands\cite{he-2016-topol-invar}, especially when they can cross each other.

In addition to inducing zeros of GFs in a single component, these factors also allow \(\det[G] = 0\) at \(\{\omega,\k\}\)s beyond the non-interacting limit\cite{wagner-2023-mott-insul}. However, we do note that our results do not exhibit such zeros near the bare band dispersions. This is because we only include static magnetic orders. To recover zeros near the bare band dispersions, we need to further include paramagnetic correlations.

Besides relating to topology, both bands of single component zeros and determinant zeros are relevant to impurity responses of Mott insulators\cite{ding-2024-local-densit}, which could also be an indicator for topology. While exact correspondence between band topology and disorder ensembles in non-interacting systems is well established\cite{chiu-2016-class-topol,altland-1997-nonst-symmet}, it is little studied in Mott systems, if not at all.

Finally, we briefly comment on the relation between zeros and Luttinger's theorem. According to \eqref{eq:d14}, both bands of zeros do not cross the chemical potential, which is fixed at \(\mu =0\) throughout this work. As discussed above, the second band of zeros is due to AFM-induced zone folding, its contribution to Luttinger's theorem is the same as the multi-band insulators. The more nontrivial contribution comes from the first type, which is due to the Hubbard-band splitting. This contribution can be reduced to either the atomic limit, or an isolated \(k\)-point. The latter perspective would be similar to the discussion with Hatsugai-Kohmoto (HK) model\cite{setty-2024-elect-proper,hatsugai-1992-exact-solvab}. For generic cases, where the AFM order is neither zero nor fully polarized, one always obtain the correct Luttinger count.
\section{Discussion}
\label{sec:orgc458899}

In this study, we establish a comprehensive theoretical framework to study the Antiferromagnetic Mott (AFM) states within the Hubbard model. We analytically derived the single-particle Green's functions (GFs) for the AFM Mott phase of the Hubbard model. Through the analysis of the GFs, we accounted for various important aspects of AFM Mott insulators which were established both numerically and experimentally. The Mott gap function explained the so called magnetic blueshift of the Mott gap. The spectral functions matches recent angle resolved photon emission (ARPES) experiments on the parental compounds of the cuprate high \(T_c\) superconductors. The remarkable alignment between our analytic theory and observations by both numerics and experiments proves that our comprehension of AFM Mott insulators is on the right track, and more importantly, provides a solid ground for further investigations on the doped AFM Mott insulators. Such alignment also demonstrates the validity and efficiency of the recently proposed \emph{algebraic dynamical theory} framework\cite{Ding-2022-algeb-dynam}, which underlies the construction of this work.

The analytic expression of the single-particle GFs for AFM Mott insulators unambiguously demonstrates previously proposed important characters of the large-\(U\) Hubbard model even away from half-filling. First, the expression clearly possesses nontrivial Luttinger surfaces or exact zeros, in addition to the atomic limit ones at \(\omega = \si m_z U/2\). The nontrivial ones are given by the solutions of \(\omega^2 -U^2/4 - \epsilon_{\k} \omega =0\), which clearly disperses. Secondly, the spectral weights vary strongly with both \(\omega\) and \(\k\). While this is observed by ARPES for a long time, theoretical understanding for such behavior is still difficult, especially for microscopic theories. It is found  previously in ECFL\cite{mai-2018-extrem-correl} calculations for the 2D \(t-J\) model and in the unrealistic Hatsugai-Kohmoto model\cite{hatsugai-1992-exact-solvab}. Finally, the explicit Hubbard band dispersion make it possible to study topological states of large-\(U\) Hubbard models more precisely.

\textcolor{black}{
While the method of this work demonstrate advantages over traditional slave particle techniques, especially in treating the spin-charge coupling issue, we briefly discuss the current insufficiencies here.
First, we are not yet able to treat the bandwidth driven Mott transition, which is also related to the debated nature of the crossover from a spin-density-wave AFM Slater insulator to a Mott AFM state\cite{hirsch-1985-two-dimen,hirsch-1989-antif-two,pruschke-2003-from-slater}. This is mostly due to the approximations we made in this work. To account for Mott transition, we need to study the spatial many-body fluctuations, noting that single electron excitations already include part of the spatial fluctuations. We refrain from saying charge fluctuations here because the transition (or crossover) is inevitably affected by magnetic fluctuations. This can be evident by tuning magnetic fluctuations by other dimensions, such as the spin size\cite{wang-2019-slater-mott}. Secondly, we start from a two-fold degenerate \(\text{N\'{e}el}\) state, which would require a degenerate perturbation theory according to quantum mechanics. However, as we know that AFM order nonetheless exists, and is not too far away from the \(\text{N\'{e}el}\) state , the current results should remain valid and compatible with that of a degenerate perturbation theory.
}
\textcolor{black}{
Thirdly, the current results are at Hartee-Fock level, if compared with traditional many-body theory. The validity of this Hartree-Fock-like theory relies on the existence of the \(\text{N\'{e}el}\)-state type of magnetic order, giving rise to a direct product-state wavefunction. Quantum fluctuations beyond the Hartree-Fock level can be naturally considered by including higher order corrections involving low-energy processes within our dynamical perturbation approach, which will be one major focus of future studies. Meanwhile, our method can be directly generalized for studying commensurate orders, provided a semi-classical state similar to the \(\text{N\'{e}el}\) state is available.
}

In conclusion, our work represents a substantial step forward in the theoretical treatment of AFM Mott states and laying a solid foundation for further exploration of the mechanism of high \(T_c\) superconductivity as doped AFM Mott insulators via the current framework.

\emph{Acknowledgement.}--- W.D. was supported by the National Key R\&D Program of the MOST of China under Grant No. 2022YFA1602603 and the Startup Grant No. S020118002/002 of Anhui University. R.Y. was supported by the National Science Foundation of China under Grant No. 12334008 and 12174441.
\bibliography{~/PhyDir/org/notes,~/PhyDir/Projects/library,~/PhyDir/org/research/refs}
\end{document}